\documentclass[aps,prl,reprint,superscriptaddress]{revtex4-2}

\usepackage{booktabs}
\usepackage{tabularx}
\usepackage{array}
\usepackage{graphicx}
\usepackage{amsmath}

\begin{document}

\title{Coordination-Driven Classification and Energetic Scaling of Boron Fullerenes and Borophenes}

\author{Nevill Gonzalez Szwacki}
\affiliation{Faculty of Physics, University of Warsaw, Pasteura 5, Warsaw, PL-02093, Poland}
\begin{abstract}
We present a comprehensive first-principles investigation of boron fullerenes and two-dimensional boron sheets, unified under a coordination-based framework. By classifying over a dozen boron nanostructures, including B$_{12}$, B$_{40}$, B$_{38}$, B$_{65}$, B$_{80}$, and B$_{92}$, according to their local atomic environments (4-, 5-, and 6-fold coordination), we identify clear trends in structural stability, electronic properties, and magnetism. A universal energetic scaling relation $E_c(n) = -a/n^b + E_c^{\mathrm{sheet}}$, with $b = 0.4$ or $b = 0.9$ depending on the coordination family, captures the convergence of fullerene cohesive energies toward those of 2D boron phases. Notably, we establish one-to-one structural correspondences between select cages and experimentally accessible borophenes: B$_{40}$ mirrors the $\chi^3$-sheet, B$_{65}$ the $\beta_{12}$-sheet, B$_{80}$ the $\alpha$-sheet, and B$_{92}$ the $bt$-sheet. Our analysis reveals two distinct families of boron nanostructures based on the scaling exponent: one with $b = 0.9$, comprising structures with significant 6-fold coordination, and another with $b = 0.4$, which includes B$_{40}$, B$_{38}$, and two experimentally observed borophenes. These clusters also exhibit large HOMO--LUMO gaps (e.g., $E_g = 1.78$~eV for B$_{40}$, 1.14~eV for B$_{92}$), contrasting with the metallicity of their 2D counterparts and, in the case of B$_{65}$, spontaneous spin polarization ($M = 3\mu_B$). Our findings provide a predictive strategy for designing boron nanostructures by leveraging coordination fingerprints, and are further validated by the recent experimental synthesis of the B$_{80}$ cage. This work bridges zero- and two-dimensional boron chemistry, offering a roadmap for the future synthesis and application of boron-based materials.
\end{abstract}

\maketitle


\begin{table*}
\caption{Summary of structural and electronic properties of boron clusters and sheets. $N$ is the number of atoms in the unit cell, $n_4$--$n_6$ are coordination counts, $E_c$ is the cohesive energy per atom (in eV), $E_g$ refers to HOMO--LUMO or band gap (in eV), $M$ is the total magnetic moment (in $\mu_B$), $d_\text{min}$ and $d_\text{max}$ are the minimum and maximum B--B bond lengths (in {\AA}), Sym. denotes point or space symmetry, and Ref. indicates the primary literature source for the particular structure.}
\begin{tabular}{lccccccccccc}
\toprule
Structure & $N$ & $n_4$ & $n_5$ & $n_6$ & $E_c$ & $E_g$ & $M$ & $d_\text{min}$ & $d_\text{max}$ & Sym. & Ref. \\
\toprule
\multicolumn{12}{c}{\textbf{5-coordinated structures}} \\
B$_{24}$ & 24 & 0 & 24 & 0 & 5.290 & 0.31 & 2 & 1.675 & 1.716 & $O$ & This work \\
B$_{60}$ & 60 & 0 & 60 & 0 & 5.671 & 0.13 & 2.11 & 1.655 & 1.747 & $I$ & \cite{Zope2011} \\
snub-sheet & 6 & 0 & 6 & 0 & 5.971 & metal & 0 & 1.676 & 1.709 & $P6/m$ & \cite{Zope2011} \\
\midrule
\multicolumn{12}{c}{\textbf{5,6-coordinated structures}} \\
B$_{16}$ & 16 & 0 & 12 & 4 & 5.089 & 0.93 & 2 & 1.658 & 1.859 & $T_d$ & This work \\
B$_{32}$ & 32 & 0 & 24 & 8 & 5.481 & 1.32 & 0 & 1.643 & 1.773 & $O_h$ & This work \\
B$_{80}$ & 80 & 0 & 60 & 20 & 5.813 & 1.02 & 0 & 1.666 & 1.737 & $I_h$ & \cite{GonzalezSzwacki2007} \\
$\alpha$-sheet & 8 & 0 & 6 & 2 & 5.991 & metal & 0 & 1.679 & 1.702 & $P6/mmm$ & \cite{Tang2007} \\
\midrule
\multicolumn{12}{c}{\textbf{5,6-coordinated structures (Boustani structures)}} \\
B$_{12}$ & 12 & 0 & 12 & 0 & 5.089 & 0.17 & 2 & 1.694 & 1.709 & $I_h$ & \cite{Boustani1997} \\
B$_{42}$ & 42 & 0 & 12 & 30 & 5.617 & 0.24 & 2.03 & 1.682 & 1.760 & $I_h$ & This work \\
B$_{92}$ & 92 & 0 & 12 & 80 & 5.781 & 1.14 & 0 & 1.632 & 1.809 & $I_h$ & \cite{GonzalezSzwacki2007} \\
$bt$-sheet & 2 & 0 & 0 & 2 & 5.906 & metal & 0 & 1.603 & 1.890 & $Pmmn$ & \cite{Boustani1999} \\
\midrule
\multicolumn{12}{c}{\textbf{4,5-coordinated structures}} \\
B$_{40}$ & 40 & 16 & 24 & 0 & 5.740 & 1.78 & 0 & 1.594 & 1.842 & $D_{2d}$ & \cite{Zhai2014} \\
$\chi_3$-sheet & 8 & 4 & 4 & 0 & 5.949 & metal & 0 & 1.613 & 1.718 & $Cmmm$ & \cite{Wu2012} \\
\midrule
\multicolumn{12}{c}{\textbf{4,5,6-coordinated structures}} \\
B$_{38}$ & 38 & 8 & 20 & 10 & 5.703 & 1.12 & 0 & 1.636 & 1.759 & $D_{2h}$ & \cite{Lv2014} \\
B$_{65}$ & 65 & 10 & 40 & 15 & 5.747 & 0.17 & 3 & 1.630 & 1.742 & $D_{5h}$ & \cite{GonzalezSzwacki2007} \\
$\beta_{12}$-sheet & 5 & 2 & 2 & 1 & 5.935 & metal & 0 & 1.642 & 1.723 & $Pmmm$ & \cite{Zhang2015} \\
\bottomrule
\end{tabular}
\label{tab:1}
\end{table*}

Motivated by the exceptional stability and electronic properties of carbon fullerenes, the focus of many studies soon turned to other elements in the periodic table in search of analogous nanostructures. Among these, boron--a neighbor of carbon with only three valence electrons--emerged as an up-and-coming candidate due to its capacity to form delocalized multi-center bonds~\cite{Boustani1997}. This unique bonding flexibility enables boron to adopt diverse coordination environments, resulting in an extraordinary variety of clusters, sheets, and extended nanostructures with no direct analogues in carbon-based chemistry~\cite{Gribanova2020}. Early theoretical efforts sought to identify energetically favorable boron cages by combining chemical intuition with computational optimization techniques. A pivotal advance came with Boustani’s Aufbau principle~\cite{Boustani1997}, which provided a systematic framework for constructing stable boron clusters through the sequential addition of specific \textit{building blocks}, typically including pentagonal pyramids (B$_6$) and hexagonal pyramids (B$_7$). Building on this approach--with a key modification involving the exclusive use of triangular and pentagonal units--led to the landmark prediction of the B$_{80}$ fullerene~\cite{GonzalezSzwacki2007}. This hollow boron cage, structurally analogous to C$_{60}$ but incorporating 20 additional atoms at the centers of the hexagons, represented a turning point in the field. The B$_{80}$ structure not only exhibits high symmetry ($I_h$) and substantial electronic stability but also introduces a novel motif based on six interwoven double rings (DRs), in contrast to the pentagon-hexagon framework characteristic of carbon fullerenes.

Subsequent studies expanded the family of boron fullerenes using several different strategies. Zope \textit{et al.}~\cite{Zope2009} proposed a systematic 80$n^2$ series of icosahedral boron cages, extending the logic of carbon’s 60$n^2$ fullerenes. These structures were generated by filling and decorating hexagonal sites in boron sheets to achieve configurations consistent with the boron $\alpha$-sheet, which had been identified as an exceptionally stable borophene phase~\cite{Tang2007}. Their stability was analyzed using the concept of hole density ($\eta$), a measure of the ratio of electron-deficient sites in a hexagonal boron network, with $\eta = 1/9$ corresponding to the most stable known boron sheet. To better understand and organize this expanding zoo of boron cages, Zope and Baruah~\cite{Zope2011} introduced a structural classification based on how boron atoms occupy the polygonal faces of the cage. They identified four primary types: (i) direct analogues of carbon fullerenes with one-to-one atom substitution, (ii) cages with boron atoms added to all polygon centers, (iii) cages with filled hexagons but hollow pentagons, and (iv) structures consistent with the motif of the boron $\alpha$-sheet. They also introduced a chiral family of boron cages--the 60$n^2$ snub fullerenes--constructed by applying snub tiling operations to hexagonal boron sheets. An alternative path was taken by Gonzalez Szwacki~\cite{GonzalezSzwacki2008}, who proposed a sequence of spherical fullerenes (B$_{12}$, B$_{80}$, B$_{180}$, B$_{300}$, etc.) derived from six interlocking DRs of increasing size. This family maintained high symmetry and showed a trend toward increasing stability with cage size, with B$_{180}$ emerging as the energetically optimal member in the sequence. In parallel, Sheng \textit{et al.}~\cite{Sheng2009} introduced the so-called S-fullerenes B$_{32+8k}$, constructed from a mix of four structural motifs--hollow quadrangles, pentagons, hexagons, and filled hexagons--and derived systematically from carbon cages via a modified leapfrog transformation. Their classification, grounded in geometric tiling and topological rules such as Euler’s theorem, highlighted the recurring role of B$_{7}$ as a basic building unit in boron nanostructures.

Despite considerable efforts, previous classification schemes for boron fullerenes have remained limited in scope--typically focusing on specific size ranges, structural motifs, or transformation rules. A universal framework capable of meaningfully comparing boron fullerenes across different sizes, topologies, and bonding patterns has yet to be established. The recent experimental validation of two distinct boron fullerenes, B$_{40}$~\cite{Zhai2014} and, more recently, B$_{80}$~\cite{Choi2024}, underscores the urgency of developing a chemically grounded and comprehensive taxonomy. While existing classifications have yielded valuable insights within individual structural families, they lack the generality needed to encompass the full spectrum of known and theoretically predicted boron cages. Moreover, they fall short of establishing clear connections with the expanding family of experimentally realized borophenes~\cite{Mannix2015,Feng2016}. A more general framework for classifying hollow boron clusters could not only unify these diverse structural types but also help identify which of the many theoretically proposed cages--particularly those with up to 100 atoms--may be viable candidates for experimental realization. This work is motivated by the need for such a general organizing principle. In the following, we examine whether atomic coordination types present in each structure can serve as a unifying descriptor across zero- and two-dimensional boron materials.


The present model is based on the premise that local atomic coordination serves as a chemically meaningful descriptor in electron-deficient systems, such as boron, where conventional $sp^2$ or $sp^3$ hybridization schemes fail to capture bonding behavior fully~\cite {Boustani1997}. By categorizing each atom within a given cluster as 4-, 5-, or 6-fold coordinated, we establish a coordination-based classification scheme applicable to both finite cages and two-dimensional boron sheets. This approach enables the systematic comparison of structurally diverse fullerenes according to a common, physically interpretable metric. The set of boron hollow cages that was used to establish a new structural classification of these structures is listed in Table~\ref{tab:1}. Most of the structures listed there have been previously reported in the literature, experimentally (B$_{40}$ and B$_{80}$) and theoretically (B$_{12}$, B$_{38}$, B$_{60}$, B$_{65}$, B$_{80}$, and B$_{92}$). The B$_{12}$ cluster is the building block of all known bulk boron phases. The B$_{38}$ cluster is a fullerene-like structure featuring four hexagons and 56 triangles in a highly symmetric cage, and it exhibits notable energetic and chemical stability, including a large HOMO--LUMO gap and double aromaticity. The B$_{60}$ cage adopts the geometry of the snub dodecahedron. B$_{65}$ is based on a $D_{5h}$-symmetric C$_{50}$ carbon fullerene scaffold reinforced by atoms atop hexagons, and B$_{92}$ is obtained by extending the well-known B$_{80}$ cage through the addition of atoms on pentagonal faces. The list is complemented with four additional structures. Based on our previous experience~\cite{GonzalezSzwacki2007} in the construction of B$_{80}$ from a truncated icosahedron (B$_{60}$) reinforced with 20 atoms positioned above hexagonal faces, we explored other Archimedean solids as structural templates for hollow boron cages. This search aimed to identify representative structures suitable for our classification framework rather than exhaustively searching for all possible clusters. To obtain the cages, additional boron atoms were placed on select polygonal faces--hexagons or pentagons--to reinforce the cages and emulate bonding patterns observed in the rest of the structures. The B$_{16}$ cluster is derived from a truncated tetrahedron with four capping atoms on hexagons, while B$_{32}$ and B$_{42}$ originate from a truncated octahedron (with eight capping atoms on hexagons) and an icosidodecahedron (with twelve capping atoms on pentagons), respectively. Finally, the B$_{24}$ cluster adopts the geometry of the snub cube. The full list of structures considered, including their symmetry, coordination environments, and references to prior studies, is summarized in Table~\ref{tab:1}. To provide a unified energetic comparison across clusters of different sizes, we employ a size-dependent scaling law for the cohesive energy per atom of the form
\[
E_c(n) = -\frac{a}{n^b} + E_c^{sheet},
\]
where $n$ is the total number of atoms in the cluster, $a$ and $b$ are fitting parameters, and $E_c^{sheet}$ is the cohesive energy of the corresponding boron sheet. This functional form captures the convergence of cluster energetics toward the cohesive limit of extended boron sheets, with the exponent $b$ reflecting the interplay between coordination and surface effects. 

All structures were optimized using first-principles density functional theory (DFT) as implemented in the Quantum ESPRESSO package~\cite{Giannozzi2009}, with the Perdew-Burke-Ernzerhof (PBE) generalized gradient approximation, norm-conserving pseudopotentials~\cite{vanSetten2018}, and a plane-wave energy cutoff of 80~Ry. Isolated clusters were treated at the $\Gamma$-point, while periodic boron sheets were computed using an $8 \times 8 \times 1$ Monkhorst-Pack $k$-point grid and a vacuum spacing of 10~{\AA} along the non-periodic direction to prevent image interactions. Convergence thresholds of $10^{-6}$~Ry for total energy and $10^{-5}$~Ry/bohr for forces were adopted throughout.


\begin{figure}
    \centering
    \includegraphics[width=\linewidth]{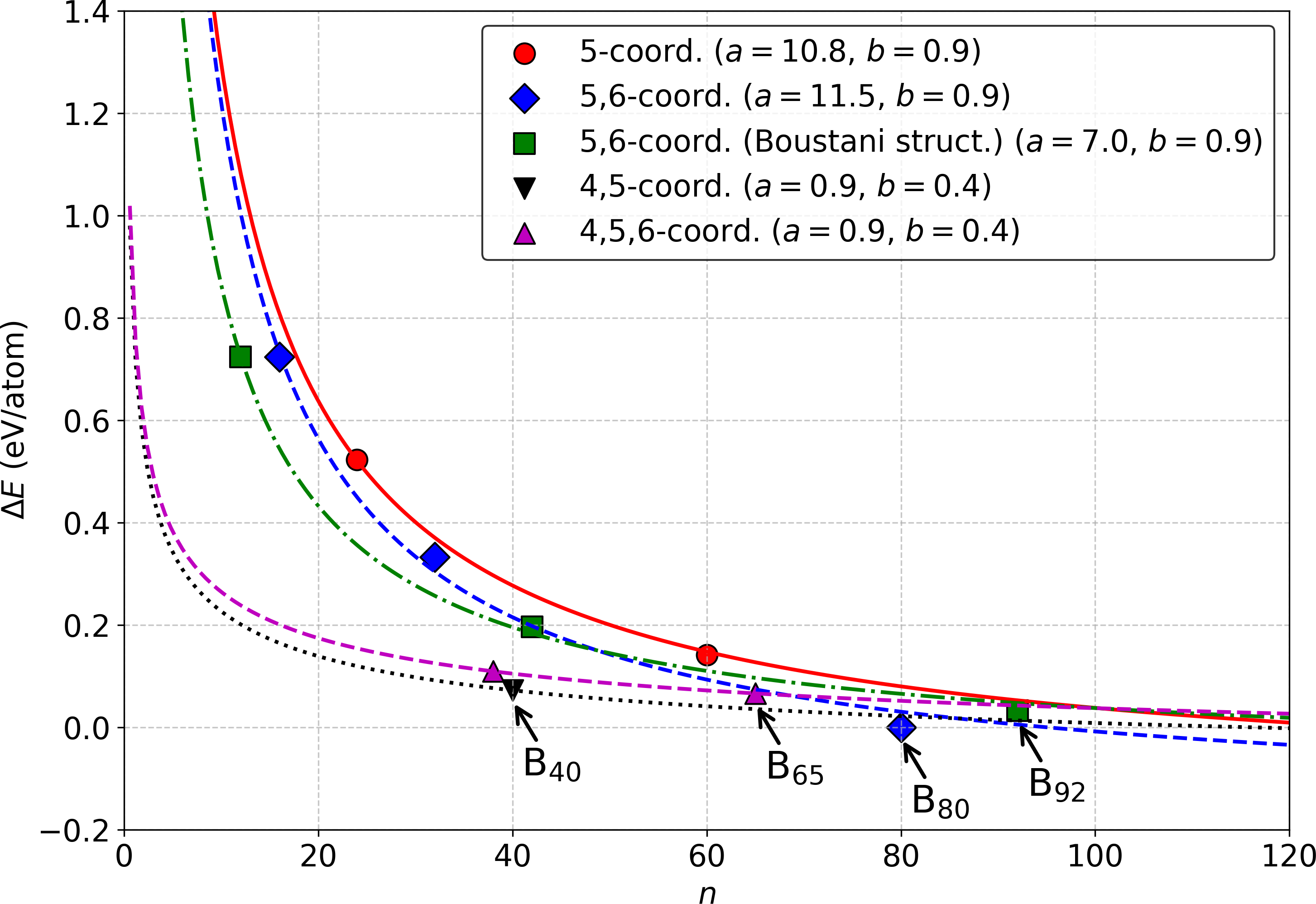}
\caption{(Color online) Relative energy per atom $\Delta E$ as a function of the number of atoms $n$ for clusters with different coordination types. Each data set corresponds to a distinct coordination configuration, indicated in the legend. Curves are fitted with the function $\Delta E(n) = E_c^{\mathrm{B}_{80}} - E_c(n) = a/n^b + c$, with fitted parameters $a$ and $b$ (or only $a$ for fixed $b$ in the case of 4,5-coordinated structures) provided in the legend, and $c = E_c^{\mathrm{B}_{80}} - E_c^{\mathrm{sheet}}$. Two distinct coordination families are identified: one characterized by $b = 0.4$ and another by $b = 0.9$. The most stable clusters, B$_{40}$, B$_{65}$, B$_{80}$, and B$_{92}$, are specifically labeled.}
    \label{fig:1}
\end{figure}

\begin{figure*}
    \centering
    \setlength{\tabcolsep}{10pt} 
    \renewcommand{\arraystretch}{1.2} 
    \begin{tabular}{c}
        \multicolumn{1}{c}{\textbf{5-coordinated atoms}} \\
        \begin{tabular}{ccc}
            \includegraphics[width=0.13\linewidth]{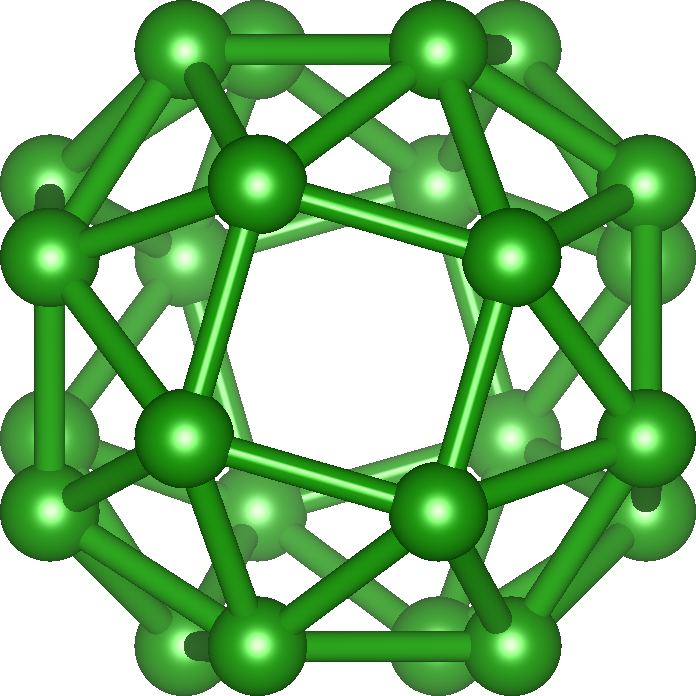} &
            \includegraphics[width=0.18\linewidth]{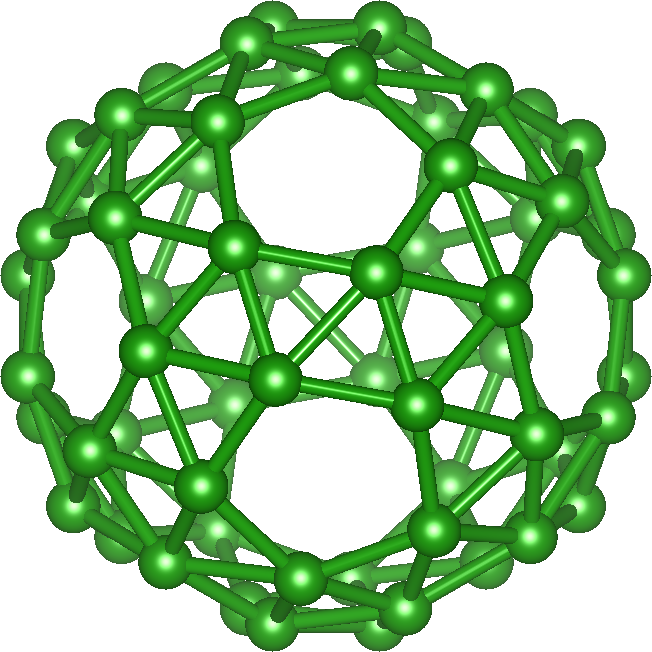} & 
            \includegraphics[width=0.28\linewidth]{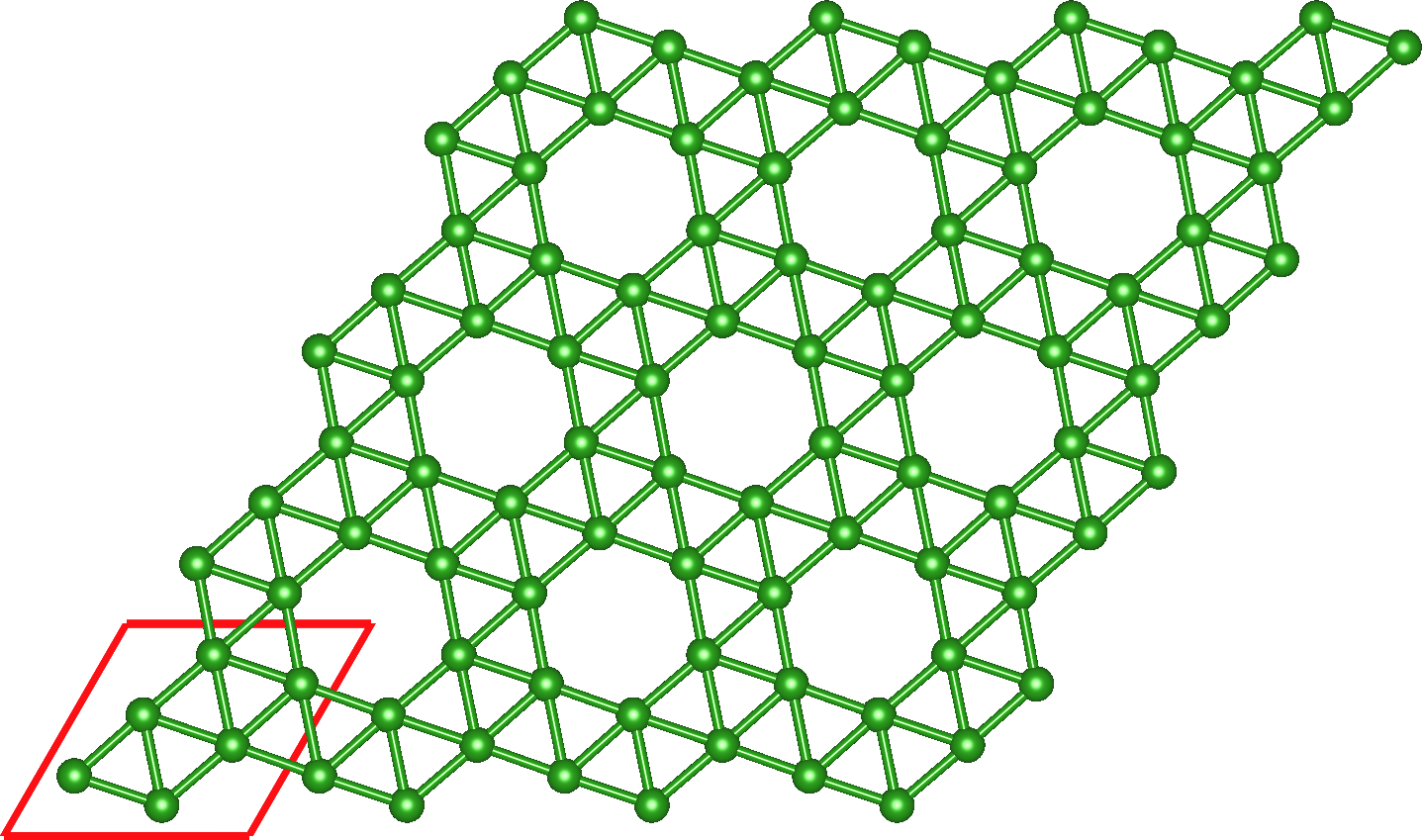} \\
            B\textsubscript{24} & B\textsubscript{60} & snub-sheet \\
        \end{tabular} \\
        [5pt]
        \multicolumn{1}{c}{\textbf{5,6-coordinated atoms}} \\
        \begin{tabular}{cccc}
            \includegraphics[width=0.14\linewidth]{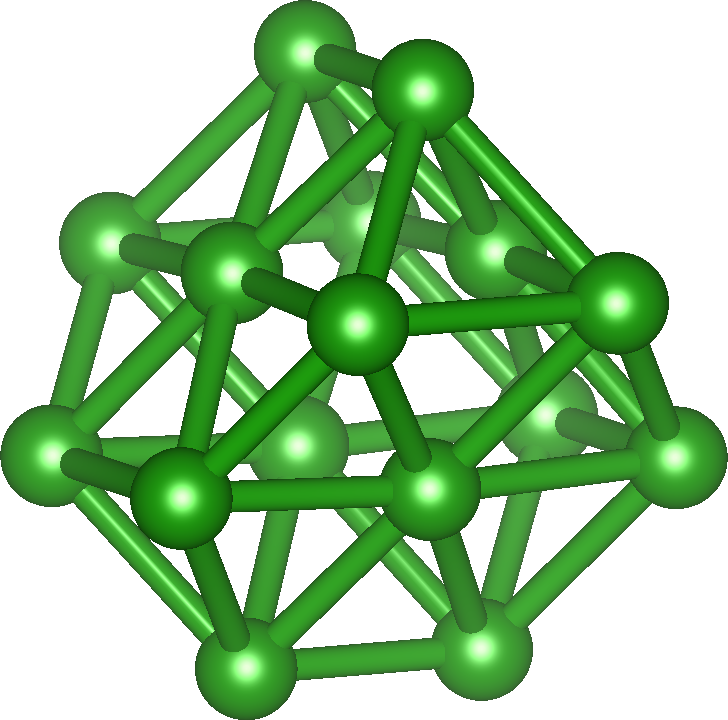} &
            \includegraphics[width=0.16\linewidth]{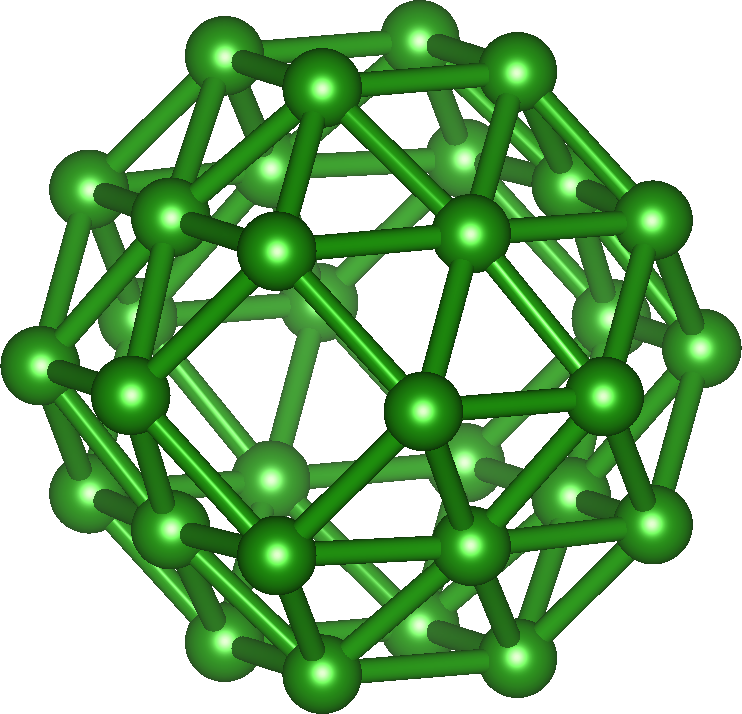} & 
            \includegraphics[width=0.16\linewidth]{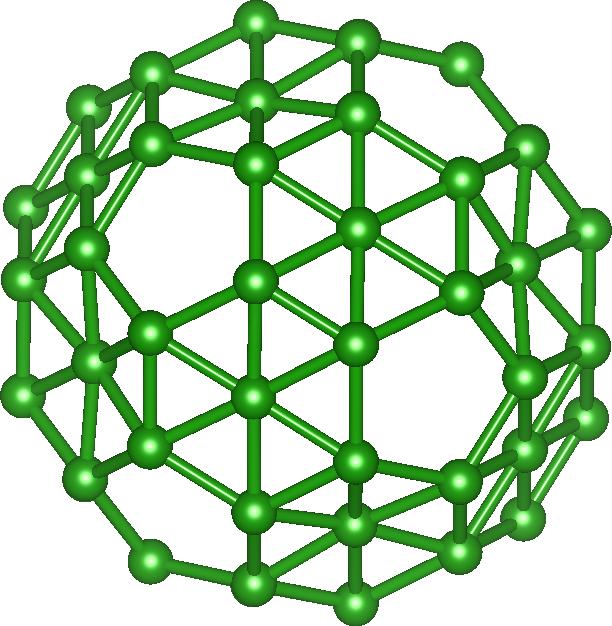} &
            \includegraphics[width=0.28\linewidth]{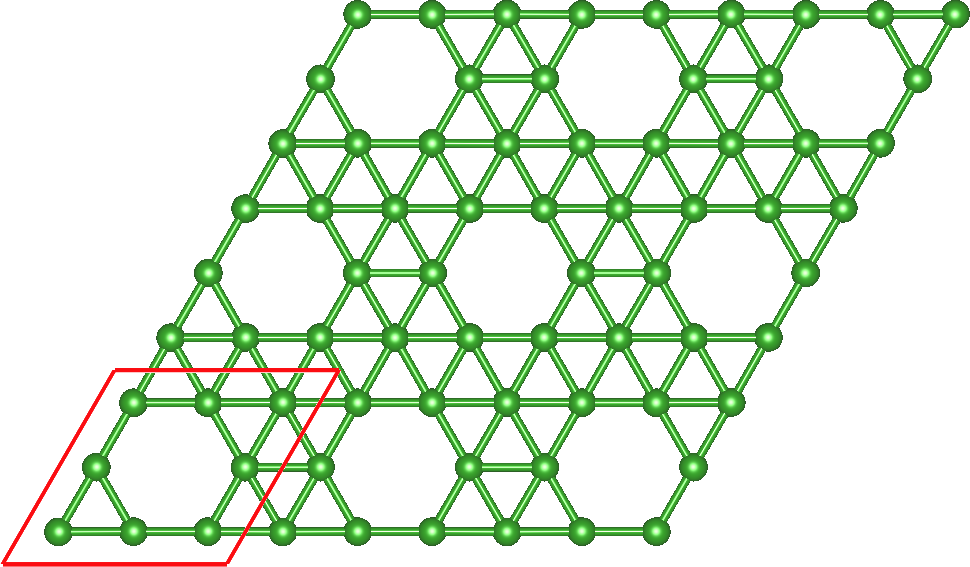} \\
            B\textsubscript{16} & B\textsubscript{32} & B\textsubscript{80} & $\alpha$-sheet \\
        \end{tabular} \\
        [5pt]
        \multicolumn{1}{c}{\textbf{5,6-coordinated atoms (Boustani cages)}} \\
        \begin{tabular}{cccc}
            \includegraphics[width=0.09\linewidth]{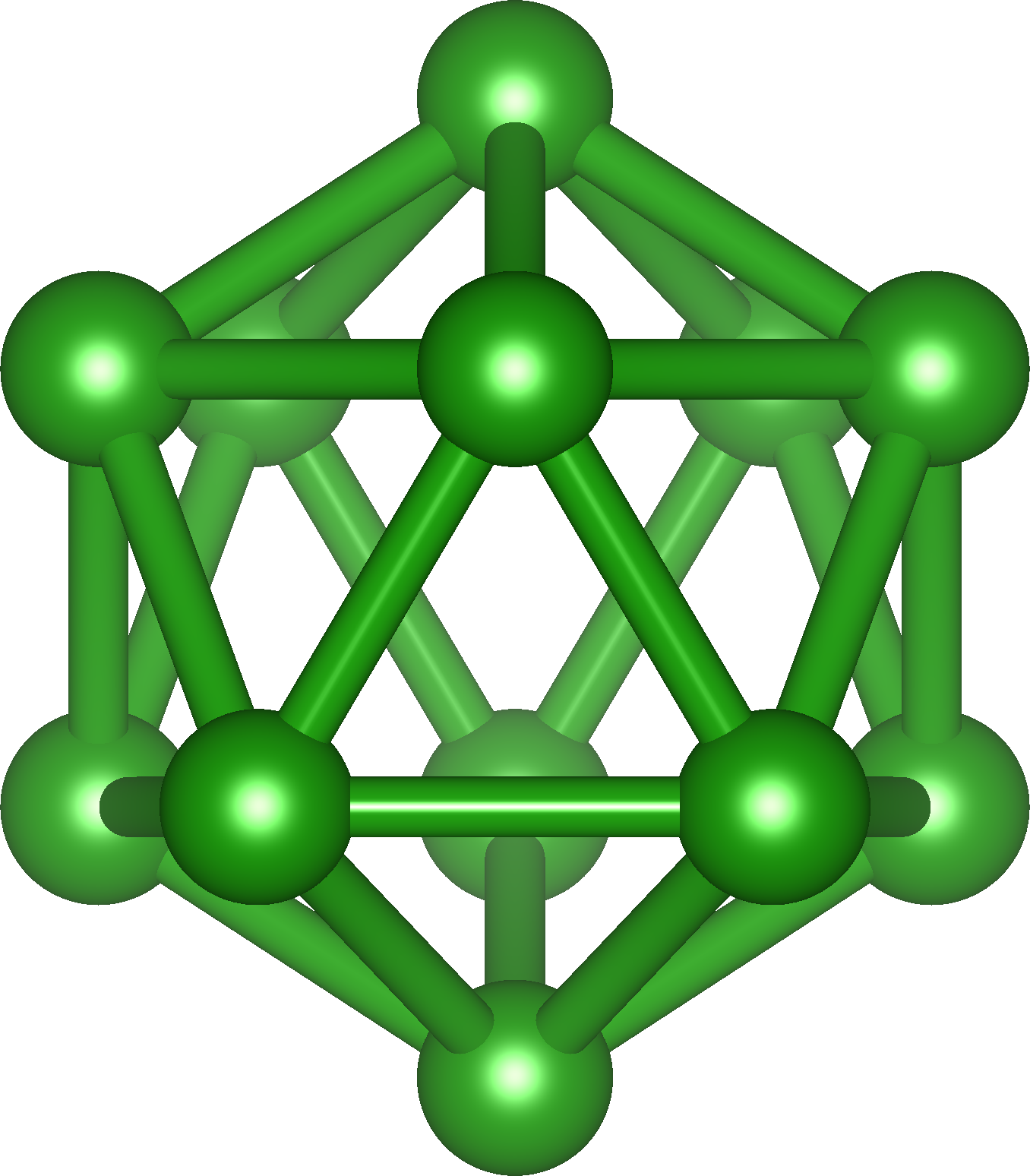} &
            \includegraphics[width=0.16\linewidth]{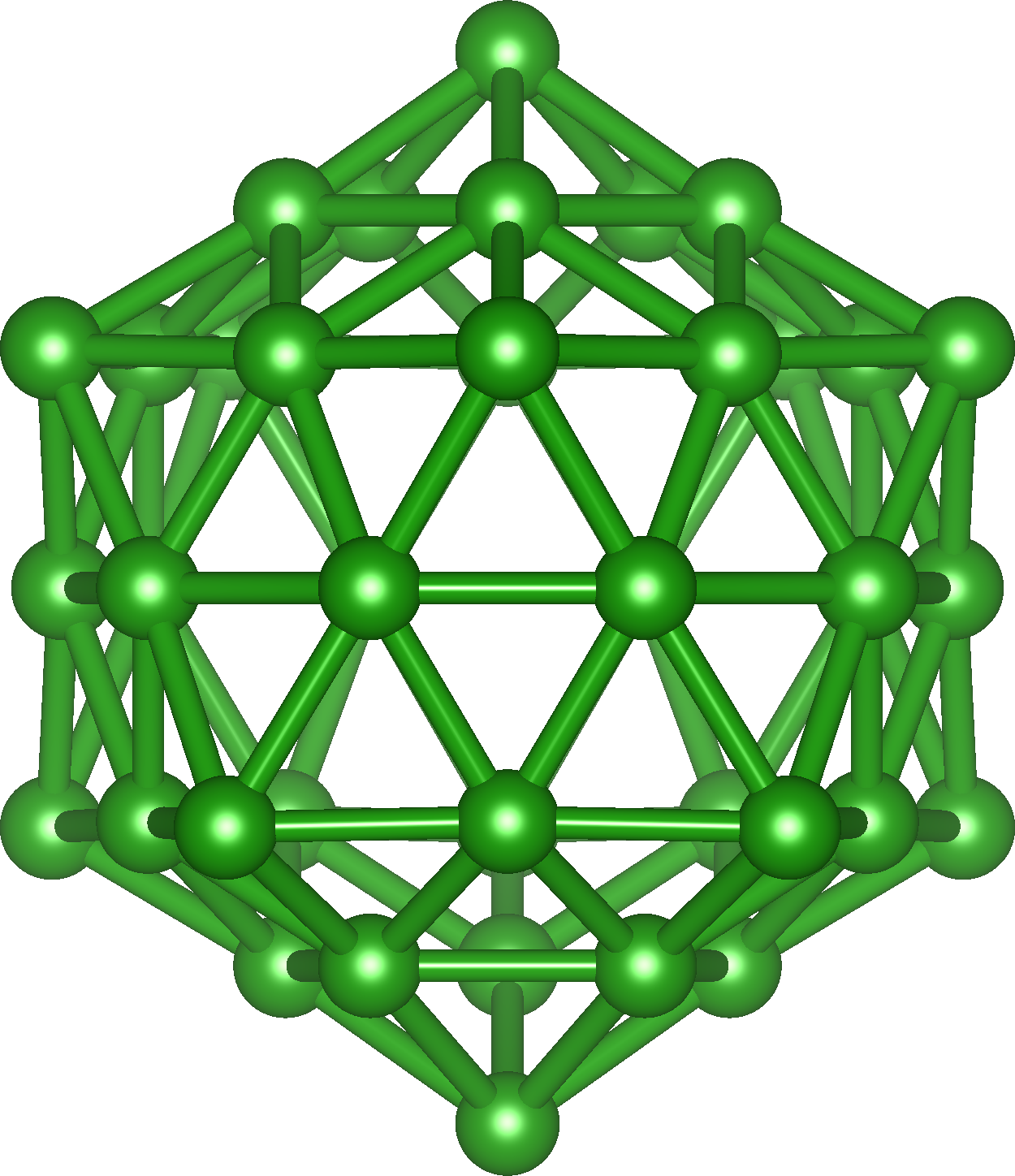} &         
            \includegraphics[width=0.17\linewidth]{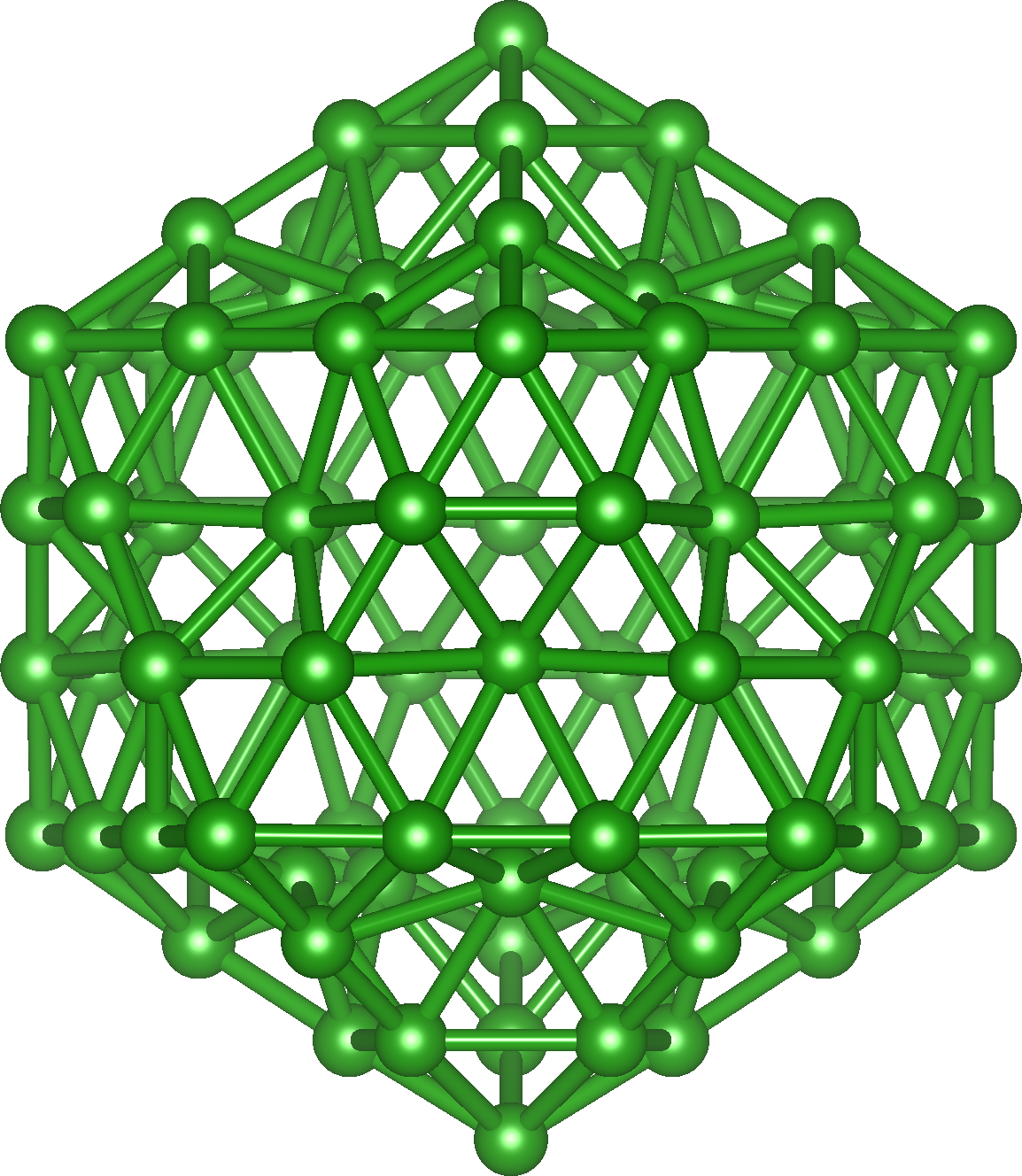} &
            \includegraphics[width=0.19\linewidth]{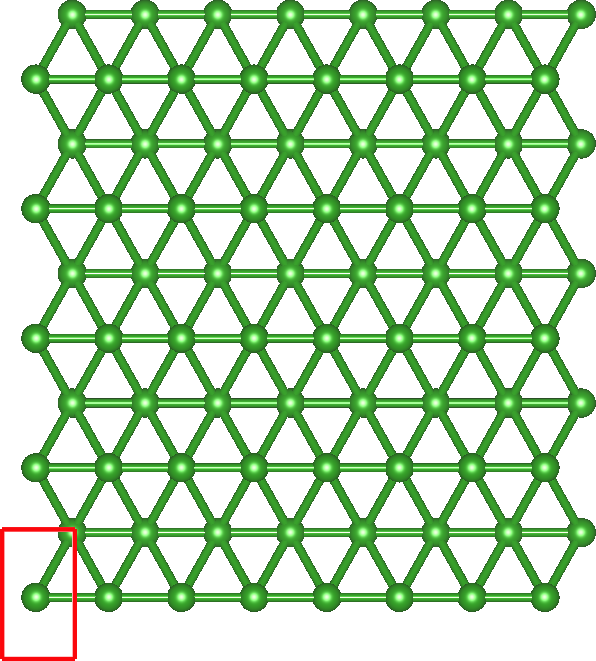} \\
            B\textsubscript{12} & B\textsubscript{42} &  B\textsubscript{92} & $bt$-sheet \\
        \end{tabular} \\
        [5pt]
        \multicolumn{1}{c}{\textbf{4,5,6-coordinated structures}} \\
        \begin{tabular}{ccc}
            \includegraphics[width=0.10\linewidth]{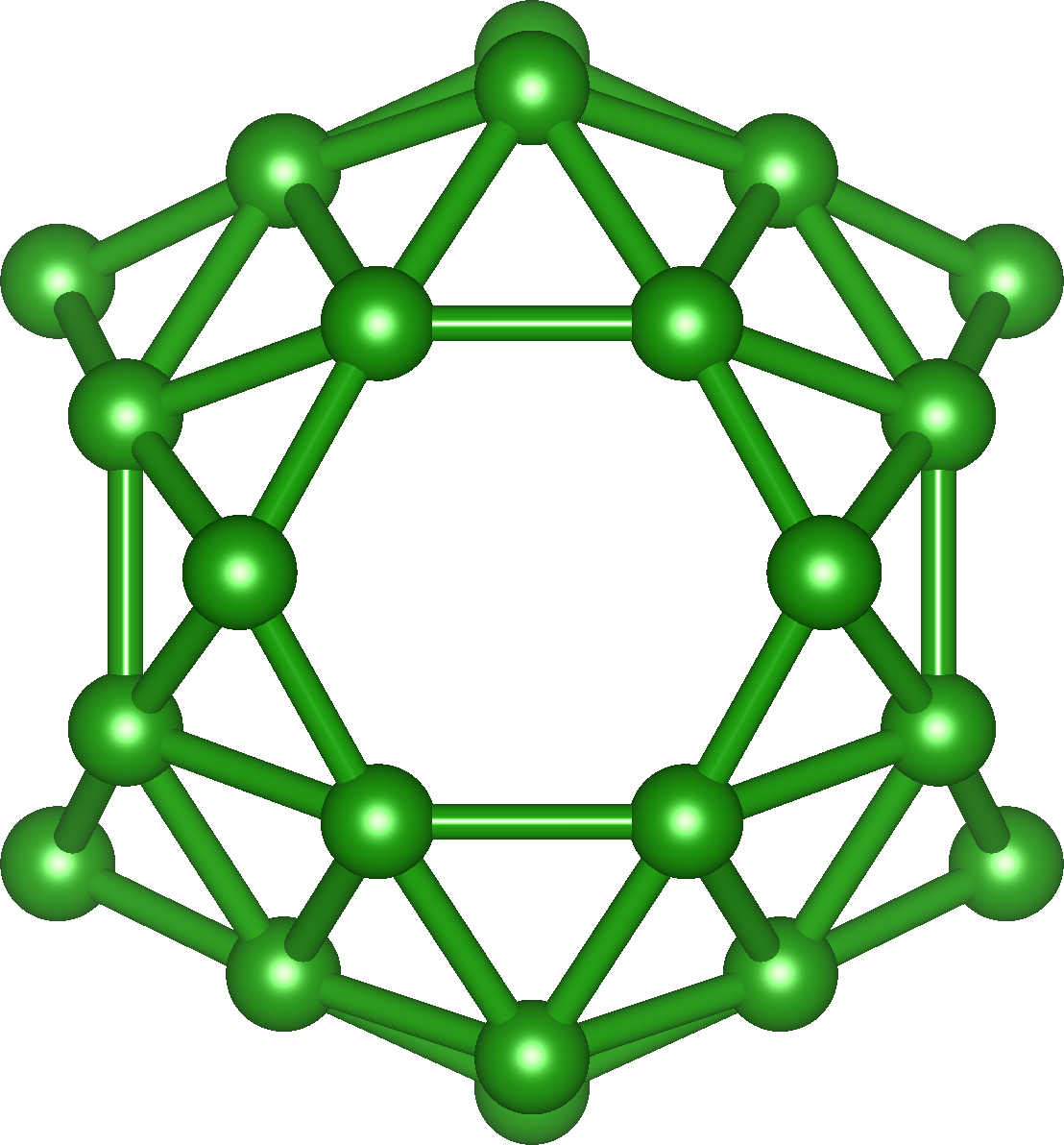} &         \includegraphics[width=0.14\linewidth]{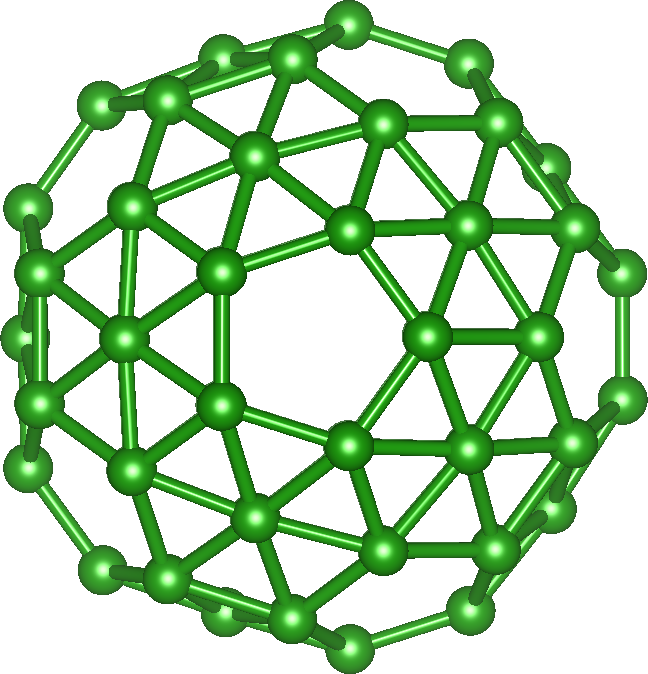} &
        \includegraphics[width=0.21\linewidth]{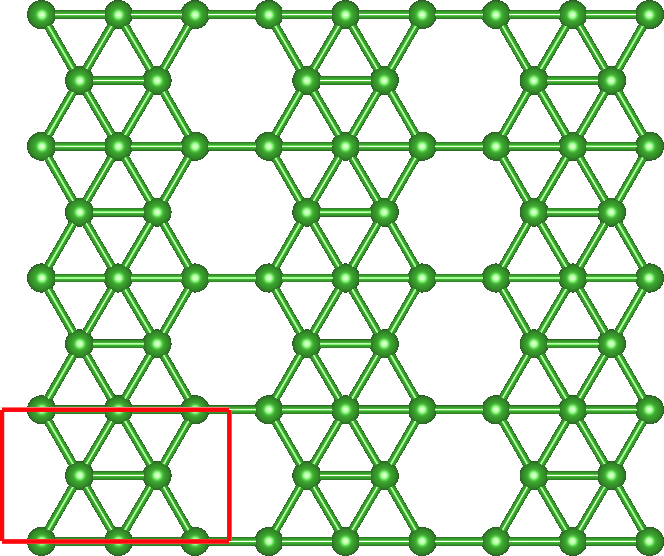} \\
            B\textsubscript{38} & B\textsubscript{65} & $\beta_{12}$-sheet \\
        \end{tabular} \\
        [5pt]
        \multicolumn{1}{c}{\textbf{4,5-coordinated structures}} \\
        \begin{tabular}{cc}
        \includegraphics[width=0.11\linewidth]{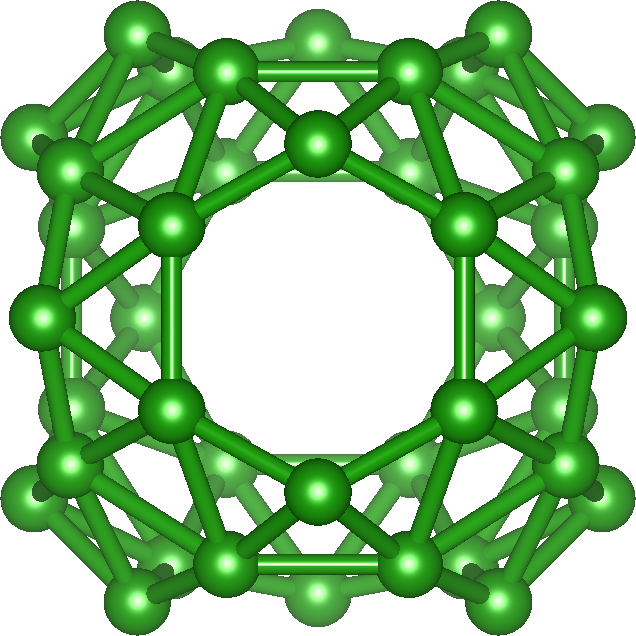} &
        \includegraphics[width=0.23\linewidth]{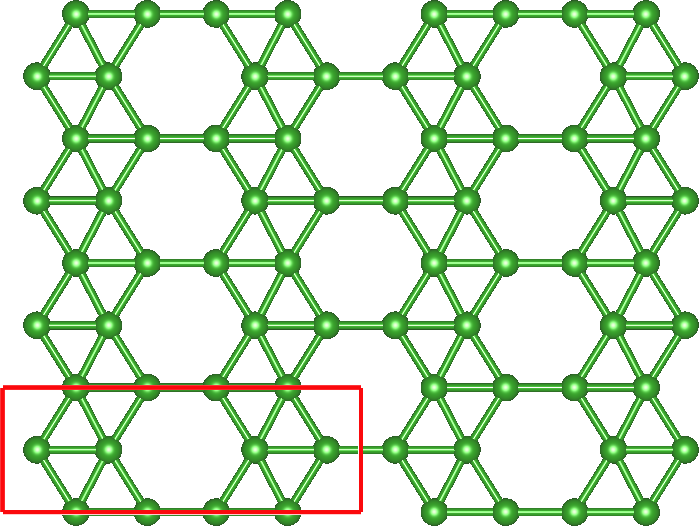} \\
        B\textsubscript{40} & $\chi_{3}$-sheet \\
        \end{tabular}
    \end{tabular}
    \caption{(Color online) Optimized geometries and structural relationships of selected boron clusters and boron sheets, grouped by dominant atomic coordination types. The representative clusters are classified according to their local coordination: 5-fold and mixed 5,6-fold, Boustani-type (5,6-fold), 4,5-fold, and 4,5,6-fold motifs.}
\label{fig:2}
\end{figure*}

This coordination-based framework enables a consistent structural classification of boron fullerenes and provides a direct link to the local bonding motifs found in experimentally and theoretically known borophene sheets. Its predictive value lies in establishing an organizational principle that bridges the structural diversity of boron clusters with the topology of extended 2D phases. The local atomic coordination environment strongly modulates the relative energetic stability of boron fullerenes. Figure~\ref{fig:1} displays the energy difference per atom, $\Delta E$, relative to the cohesive energy of B$_{80}$ ($E_c^{\mathrm{B}_{80}}$), plotted as a function of the total number of atoms $n$. Structures are grouped according to dominant coordination motifs—purely 5-fold, mixed 5,6-fold, hybrid 4,5,6-fold, and so forth—and fitted to the empirical scaling law $\Delta E(n) = a/n^b + c$. The fitted exponent $b$ distinguishes two coordination families: one with $b = 0.9$ for clusters dominated by 5- and 5,6-fold coordination, and another with $b = 0.4$ for structures involving 4,5- and 4,5,6-fold motifs. This approach captures the general trend that strain energy per atom decreases with increasing cluster size, as larger fullerenes approximate a spherical geometry and reduce distortions from planar bonding preferences. The most stable fullerenes within each coordination family consistently exhibit a higher degree of atomic coordination. For instance, B$_{80}$ and B$_{92}$ (with 20 and 80 atoms in 6-fold coordination, respectively) have cohesive energies of 5.813~eV and 5.781~eV per atom, among the highest in the set. B$_{92}$ is derived from B$_{80}$ by capping its pentagonal faces with twelve additional atoms. At the same time, it features more 6-fold coordinated atoms, B$_{80}$ remains slightly more stable due to a better balance between 5- and 6-fold coordinated sites. In contrast, purely 5-coordinated structures such as B$_{24}$ and B$_{60}$ are less favorable, with $E_c = 5.290$~eV and 5.671~eV, respectively, despite their high symmetry ($O$ and $I$). Structures with mixed coordination, such as B$_{40}$ ($n_4 = 16$, $n_5 = 24$), B$_{38}$ ($n_4 = 8$, $n_5 = 20$, $n_6 = 10$), and B$_{65}$ ($n_4 = 10$, $n_5 = 40$, $n_6 = 15$), display enhanced energetic stability ($E_c = 5.740$~eV, 5.703~eV, and 5.747~eV, respectively), highlighting the advantage of combining multiple coordination environments. These trends support the conclusion that coordination number distributions are predictive of fullerene stability, providing a quantitative metric for comparing structurally diverse boron nanoclusters.


Visual analysis of optimized cage geometries (Figure~\ref{fig:2}) reveals that local coordination motifs not only determine energetic preferences but also provide a structural bridge between 0D fullerenes and 2D boron sheets. B$_{24}$ and B$_{60}$, although purely 5-coordinated and energetically less favorable despite their high symmetry, are structurally related to the periodic snub-sheet, highlighting their role as finite analogues of extended 2D phases. More complex coordination patterns yield deeper structural correspondences: B$_{40}$ shares its 4,5-coordination scheme with the $\chi_3$-sheet and adopts $D_{2d}$ symmetry; B$_{65}$ reflects the atomic environment of the $\beta_{12}$-sheet with a mixed 4,5,6-fold coordination; B$_{80}$, with atoms centered on hexagons, mirrors the $\alpha$-sheet; and B$_{92}$ further extends this motif by adding twelve atoms atop the pentagons of B$_{80}$, closely resembling the densely packed $bt$-sheet. Overall, structures combining diverse coordination environments are preferred over uniformly coordinated ones--both in fullerenes and borophenes--and the observed structural analogies provide a rational framework for interpreting boron fullerenes as finite reconstructions of stable 2D boron motifs.


Table~\ref{tab:1} details the electronic gaps and total magnetic moments for all clusters and sheet models. Clusters with substantial 6-fold coordination, such as B$_{80}$ and B$_{92}$, possess large HOMO--LUMO gaps (1.02~eV and 1.14~eV, respectively), indicative of electronic stability and low chemical reactivity. Notably, B$_{40}$ exhibits the largest gap (1.78~eV) despite its lower coordination, likely due to enhanced orbital localization arising from its compact and symmetric $D_{2d}$ geometry. Other structures with mixed coordination, including B$_{38}$ and B$_{65}$, also show moderate-to-large gaps (1.12~eV and 0.17~eV, respectively), underscoring the stabilizing role of hybrid bonding motifs. In contrast, purely 5-fold coordinated clusters such as B$_{24}$ and B$_{60}$ have much smaller gaps (0.31~eV and 0.13~eV), approaching metallic behavior. These trends underscore the role of the coordination environment in shaping electronic properties, with higher and mixed coordination leading to more stable electronic configurations. Unlike the cages, their corresponding two-dimensional boron sheets are metallic under the same computational conditions, highlighting the distinct electronic character that emerges in finite versus extended systems. Magnetism arises in clusters with mixed or frustrated coordination patterns: B$_{65}$ exhibits a net magnetic moment of $3\,\mu_B$, absent in its 2D analogue; similar spin polarization is found in B$_{12}$ ($2\,\mu_B$), B$_{16}$ ($2\,\mu_B$), and B$_{42}$ ($2.03\,\mu_B$). None of the extended boron sheets—$\chi_3$, $\beta_{12}$, $\alpha$, or $bt$—exhibit magnetization, suggesting that finite size and geometric frustration are essential for inducing magnetic moments. These results demonstrate that both electronic and magnetic properties depend sensitively on local coordination and its spatial distribution, providing a predictive handle for designing boron nanostructures with targeted functionalities.


The observed trends in cohesive energy, electronic structure, and magnetism reinforce the central premise of this work: that local atomic coordination serves as a chemically meaningful and predictive descriptor for boron nanostructures. Clusters with coordination fingerprints similar to known borophene phases tend to exhibit comparable energetic and electronic behavior, while purely 5-coordinated or irregular topologies deviate from this trend. As shown in Figures~\ref{fig:1} and~\ref{fig:2}, and Table~\ref{tab:1}, this coordination-driven classification enables the evaluation of structurally and electronically diverse fullerenes on a common footing, offering a systematic framework for interpreting their stability, electronic properties, and magnetism. This approach offers a practical set of design rules for constructing new boron-based nanostructures with tailored functionalities, leveraging the predictive power of local bonding environments.


Summarizing, we have presented a comprehensive coordination-based classification of boron fullerenes, linking their energetic, electronic, and magnetic properties to their local bonding environments. Using first-principles calculations, we analyzed a representative set of cage geometries, ranging from purely 5-coordinated high-symmetry clusters to hybrid 4,5,6-coordinated cages with direct analogs in borophene sheets. Energetic stability is found to increase with coordination number, with the most stable structures--B$_{80}$ and B$_{92}$--featuring large fractions of 6-fold coordinated atoms. The relative energy scaling follows two distinct universal trends with cluster size, well described by $\Delta E(n) = a/n^b + c$ with $b = 0.4$ or $b = 0.9$, indicating convergence toward sheet-like cohesive energies depending on coordination type. Clusters with significant 6-fold coordination, such as B$_{80}$ and B$_{92}$, exhibit large HOMO--LUMO gaps (1.02 and 1.14~eV), indicating enhanced electronic stability. B$_{40}$ shows the largest gap (1.78~eV) despite lower coordination due to orbital localization from its compact $D_{2d}$ geometry. In contrast, purely 5-fold coordinated clusters like B$_{24}$ and B$_{60}$ have much smaller gaps (0.31 and 0.13~eV), approaching metallic behavior. The corresponding 2D boron sheets are metallic, underscoring the electronic differences between finite and extended systems. Our coordination-based framework also identifies one-to-one correspondences between fullerene cages and known borophene sheets. B$_{40}$ mirrors the $\chi_3$-sheet, B$_{80}$ mirrors the $\alpha$-sheet, B$_{65}$ corresponds to the $\beta_{12}$-sheet, and B$_{92}$ closely resembles the $bt$-sheet. According to this simple model, the experimentally accessible structures (B$_{40}$, $\chi_3$-sheet) and (B$_{80}$, $\alpha$-sheet) can be complemented with two new pairs--(B$_{65}$, $\beta_{12}$-sheet) and (B$_{92}$, $bt$-sheet)--for which the cage clusters should also, in principle, be experimentally feasible. These findings demonstrate that local coordination provides a unifying and predictive metric for organizing and designing boron nanostructures across dimensions. The structural and electronic coherence between fullerenes and borophenes revealed here offers a roadmap for identifying stable, functional boron materials suitable for experimental realization.

\begin{acknowledgments}
This research was funded by the National Science Centre, Poland, under grant number 2021/43/O/ST3/03280. Computational resources were provided by the Interdisciplinary Centre for Mathematical and Computational Modelling (ICM) at the University of Warsaw.
\end{acknowledgments}

\bibliographystyle{apsrev4-2}
\bibliography{v5}

\begin{thebibliography}{18}%
\makeatletter
\providecommand \@ifxundefined [1]{%
 \@ifx{#1\undefined}
}%
\providecommand \@ifnum [1]{%
 \ifnum #1\expandafter \@firstoftwo
 \else \expandafter \@secondoftwo
 \fi
}%
\providecommand \@ifx [1]{%
 \ifx #1\expandafter \@firstoftwo
 \else \expandafter \@secondoftwo
 \fi
}%
\providecommand \natexlab [1]{#1}%
\providecommand \enquote  [1]{``#1''}%
\providecommand \bibnamefont  [1]{#1}%
\providecommand \bibfnamefont [1]{#1}%
\providecommand \citenamefont [1]{#1}%
\providecommand \href@noop [0]{\@secondoftwo}%
\providecommand \href [0]{\begingroup \@sanitize@url \@href}%
\providecommand \@href[1]{\@@startlink{#1}\@@href}%
\providecommand \@@href[1]{\endgroup#1\@@endlink}%
\providecommand \@sanitize@url [0]{\catcode `\\12\catcode `\$12\catcode `\&12\catcode `\#12\catcode `\^12\catcode `\_12\catcode `\%12\relax}%
\providecommand \@@startlink[1]{}%
\providecommand \@@endlink[0]{}%
\providecommand \url  [0]{\begingroup\@sanitize@url \@url }%
\providecommand \@url [1]{\endgroup\@href {#1}{\urlprefix }}%
\providecommand \urlprefix  [0]{URL }%
\providecommand \Eprint [0]{\href }%
\providecommand \doibase [0]{https://doi.org/}%
\providecommand \selectlanguage [0]{\@gobble}%
\providecommand \bibinfo  [0]{\@secondoftwo}%
\providecommand \bibfield  [0]{\@secondoftwo}%
\providecommand \translation [1]{[#1]}%
\providecommand \BibitemOpen [0]{}%
\providecommand \bibitemStop [0]{}%
\providecommand \bibitemNoStop [0]{.\EOS\space}%
\providecommand \EOS [0]{\spacefactor3000\relax}%
\providecommand \BibitemShut  [1]{\csname bibitem#1\endcsname}%
\let\auto@bib@innerbib\@empty
\bibitem [{\citenamefont {Zope}\ and\ \citenamefont {Baruah}(2011)}]{Zope2011}%
  \BibitemOpen
  \bibfield  {author} {\bibinfo {author} {\bibfnamefont {R.~R.}\ \bibnamefont {Zope}}\ and\ \bibinfo {author} {\bibfnamefont {T.}~\bibnamefont {Baruah}},\ }\href {https://doi.org/10.1016/j.cplett.2010.11.012} {\bibfield  {journal} {\bibinfo  {journal} {Chemical Physics Letters}\ }\textbf {\bibinfo {volume} {501}},\ \bibinfo {pages} {193–196} (\bibinfo {year} {2011})}\BibitemShut {NoStop}%
\bibitem [{\citenamefont {Gonzalez~Szwacki}\ \emph {et~al.}(2007)\citenamefont {Gonzalez~Szwacki}, \citenamefont {Sadrzadeh},\ and\ \citenamefont {Yakobson}}]{GonzalezSzwacki2007}%
  \BibitemOpen
  \bibfield  {author} {\bibinfo {author} {\bibfnamefont {N.}~\bibnamefont {Gonzalez~Szwacki}}, \bibinfo {author} {\bibfnamefont {A.}~\bibnamefont {Sadrzadeh}},\ and\ \bibinfo {author} {\bibfnamefont {B.~I.}\ \bibnamefont {Yakobson}},\ }\href {https://doi.org/10.1103/physrevlett.98.166804} {\bibfield  {journal} {\bibinfo  {journal} {Physical Review Letters}\ }\textbf {\bibinfo {volume} {98}},\ \bibinfo {pages} {166804} (\bibinfo {year} {2007})}\BibitemShut {NoStop}%
\bibitem [{\citenamefont {Tang}\ and\ \citenamefont {Ismail-Beigi}(2007)}]{Tang2007}%
  \BibitemOpen
  \bibfield  {author} {\bibinfo {author} {\bibfnamefont {H.}~\bibnamefont {Tang}}\ and\ \bibinfo {author} {\bibfnamefont {S.}~\bibnamefont {Ismail-Beigi}},\ }\href {https://doi.org/10.1103/physrevlett.99.115501} {\bibfield  {journal} {\bibinfo  {journal} {Physical Review Letters}\ }\textbf {\bibinfo {volume} {99}},\ \bibinfo {pages} {115501} (\bibinfo {year} {2007})}\BibitemShut {NoStop}%
\bibitem [{\citenamefont {Boustani}(1997)}]{Boustani1997}%
  \BibitemOpen
  \bibfield  {author} {\bibinfo {author} {\bibfnamefont {I.}~\bibnamefont {Boustani}},\ }\href {https://doi.org/10.1103/physrevb.55.16426} {\bibfield  {journal} {\bibinfo  {journal} {Physical Review B}\ }\textbf {\bibinfo {volume} {55}},\ \bibinfo {pages} {16426–16438} (\bibinfo {year} {1997})}\BibitemShut {NoStop}%
\bibitem [{\citenamefont {Boustani}\ \emph {et~al.}(1999)\citenamefont {Boustani}, \citenamefont {Quandt}, \citenamefont {Hernández},\ and\ \citenamefont {Rubio}}]{Boustani1999}%
  \BibitemOpen
  \bibfield  {author} {\bibinfo {author} {\bibfnamefont {I.}~\bibnamefont {Boustani}}, \bibinfo {author} {\bibfnamefont {A.}~\bibnamefont {Quandt}}, \bibinfo {author} {\bibfnamefont {E.}~\bibnamefont {Hernández}},\ and\ \bibinfo {author} {\bibfnamefont {A.}~\bibnamefont {Rubio}},\ }\href {https://doi.org/10.1063/1.477976} {\bibfield  {journal} {\bibinfo  {journal} {The Journal of Chemical Physics}\ }\textbf {\bibinfo {volume} {110}},\ \bibinfo {pages} {3176–3185} (\bibinfo {year} {1999})}\BibitemShut {NoStop}%
\bibitem [{\citenamefont {Zhai}\ \emph {et~al.}(2014)\citenamefont {Zhai}, \citenamefont {Zhao}, \citenamefont {Li}, \citenamefont {Chen}, \citenamefont {Bai}, \citenamefont {Hu}, \citenamefont {Piazza}, \citenamefont {Tian}, \citenamefont {Lu}, \citenamefont {Wu}, \citenamefont {Mu}, \citenamefont {Wei}, \citenamefont {Liu}, \citenamefont {Li}, \citenamefont {Li},\ and\ \citenamefont {Wang}}]{Zhai2014}%
  \BibitemOpen
  \bibfield  {author} {\bibinfo {author} {\bibfnamefont {H.-J.}\ \bibnamefont {Zhai}}, \bibinfo {author} {\bibfnamefont {Y.-F.}\ \bibnamefont {Zhao}}, \bibinfo {author} {\bibfnamefont {W.-L.}\ \bibnamefont {Li}}, \bibinfo {author} {\bibfnamefont {Q.}~\bibnamefont {Chen}}, \bibinfo {author} {\bibfnamefont {H.}~\bibnamefont {Bai}}, \bibinfo {author} {\bibfnamefont {H.-S.}\ \bibnamefont {Hu}}, \bibinfo {author} {\bibfnamefont {Z.~A.}\ \bibnamefont {Piazza}}, \bibinfo {author} {\bibfnamefont {W.-J.}\ \bibnamefont {Tian}}, \bibinfo {author} {\bibfnamefont {H.-G.}\ \bibnamefont {Lu}}, \bibinfo {author} {\bibfnamefont {Y.-B.}\ \bibnamefont {Wu}}, \bibinfo {author} {\bibfnamefont {Y.-W.}\ \bibnamefont {Mu}}, \bibinfo {author} {\bibfnamefont {G.-F.}\ \bibnamefont {Wei}}, \bibinfo {author} {\bibfnamefont {Z.-P.}\ \bibnamefont {Liu}}, \bibinfo {author} {\bibfnamefont {J.}~\bibnamefont {Li}}, \bibinfo {author} {\bibfnamefont {S.-D.}\ \bibnamefont {Li}},\ and\ \bibinfo {author} {\bibfnamefont {L.-S.}\ \bibnamefont
  {Wang}},\ }\href {https://doi.org/10.1038/nchem.1999} {\bibfield  {journal} {\bibinfo  {journal} {Nature Chemistry}\ }\textbf {\bibinfo {volume} {6}},\ \bibinfo {pages} {727–731} (\bibinfo {year} {2014})}\BibitemShut {NoStop}%
\bibitem [{\citenamefont {Wu}\ \emph {et~al.}(2012)\citenamefont {Wu}, \citenamefont {Dai}, \citenamefont {Zhao}, \citenamefont {Zhuo}, \citenamefont {Yang},\ and\ \citenamefont {Zeng}}]{Wu2012}%
  \BibitemOpen
  \bibfield  {author} {\bibinfo {author} {\bibfnamefont {X.}~\bibnamefont {Wu}}, \bibinfo {author} {\bibfnamefont {J.}~\bibnamefont {Dai}}, \bibinfo {author} {\bibfnamefont {Y.}~\bibnamefont {Zhao}}, \bibinfo {author} {\bibfnamefont {Z.}~\bibnamefont {Zhuo}}, \bibinfo {author} {\bibfnamefont {J.}~\bibnamefont {Yang}},\ and\ \bibinfo {author} {\bibfnamefont {X.~C.}\ \bibnamefont {Zeng}},\ }\href {https://doi.org/10.1021/nn302696v} {\bibfield  {journal} {\bibinfo  {journal} {ACS Nano}\ }\textbf {\bibinfo {volume} {6}},\ \bibinfo {pages} {7443–7453} (\bibinfo {year} {2012})}\BibitemShut {NoStop}%
\bibitem [{\citenamefont {Lv}\ \emph {et~al.}(2014)\citenamefont {Lv}, \citenamefont {Wang}, \citenamefont {Zhu},\ and\ \citenamefont {Ma}}]{Lv2014}%
  \BibitemOpen
  \bibfield  {author} {\bibinfo {author} {\bibfnamefont {J.}~\bibnamefont {Lv}}, \bibinfo {author} {\bibfnamefont {Y.}~\bibnamefont {Wang}}, \bibinfo {author} {\bibfnamefont {L.}~\bibnamefont {Zhu}},\ and\ \bibinfo {author} {\bibfnamefont {Y.}~\bibnamefont {Ma}},\ }\href {https://doi.org/10.1039/c4nr01846j} {\bibfield  {journal} {\bibinfo  {journal} {Nanoscale}\ }\textbf {\bibinfo {volume} {6}},\ \bibinfo {pages} {11692–11696} (\bibinfo {year} {2014})}\BibitemShut {NoStop}%
\bibitem [{\citenamefont {Zhang}\ \emph {et~al.}(2015)\citenamefont {Zhang}, \citenamefont {Yang}, \citenamefont {Gao},\ and\ \citenamefont {Yakobson}}]{Zhang2015}%
  \BibitemOpen
  \bibfield  {author} {\bibinfo {author} {\bibfnamefont {Z.}~\bibnamefont {Zhang}}, \bibinfo {author} {\bibfnamefont {Y.}~\bibnamefont {Yang}}, \bibinfo {author} {\bibfnamefont {G.}~\bibnamefont {Gao}},\ and\ \bibinfo {author} {\bibfnamefont {B.~I.}\ \bibnamefont {Yakobson}},\ }\href {https://doi.org/10.1002/anie.201505425} {\bibfield  {journal} {\bibinfo  {journal} {Angewandte Chemie International Edition}\ }\textbf {\bibinfo {volume} {54}},\ \bibinfo {pages} {13022–13026} (\bibinfo {year} {2015})}\BibitemShut {NoStop}%
\bibitem [{\citenamefont {Gribanova}\ \emph {et~al.}(2020)\citenamefont {Gribanova}, \citenamefont {Minyaev}, \citenamefont {Minkin},\ and\ \citenamefont {Boldyrev}}]{Gribanova2020}%
  \BibitemOpen
  \bibfield  {author} {\bibinfo {author} {\bibfnamefont {T.~N.}\ \bibnamefont {Gribanova}}, \bibinfo {author} {\bibfnamefont {R.~M.}\ \bibnamefont {Minyaev}}, \bibinfo {author} {\bibfnamefont {V.~I.}\ \bibnamefont {Minkin}},\ and\ \bibinfo {author} {\bibfnamefont {A.~I.}\ \bibnamefont {Boldyrev}},\ }\href {https://doi.org/10.1007/s11224-020-01606-9} {\bibfield  {journal} {\bibinfo  {journal} {Structural Chemistry}\ }\textbf {\bibinfo {volume} {31}},\ \bibinfo {pages} {2105–2128} (\bibinfo {year} {2020})}\BibitemShut {NoStop}%
\bibitem [{\citenamefont {Zope}\ \emph {et~al.}(2009)\citenamefont {Zope}, \citenamefont {Baruah}, \citenamefont {Lau}, \citenamefont {Liu}, \citenamefont {Pederson},\ and\ \citenamefont {Dunlap}}]{Zope2009}%
  \BibitemOpen
  \bibfield  {author} {\bibinfo {author} {\bibfnamefont {R.~R.}\ \bibnamefont {Zope}}, \bibinfo {author} {\bibfnamefont {T.}~\bibnamefont {Baruah}}, \bibinfo {author} {\bibfnamefont {K.~C.}\ \bibnamefont {Lau}}, \bibinfo {author} {\bibfnamefont {A.~Y.}\ \bibnamefont {Liu}}, \bibinfo {author} {\bibfnamefont {M.~R.}\ \bibnamefont {Pederson}},\ and\ \bibinfo {author} {\bibfnamefont {B.~I.}\ \bibnamefont {Dunlap}},\ }\href {https://doi.org/10.1103/physrevb.79.161403} {\bibfield  {journal} {\bibinfo  {journal} {Physical Review B}\ }\textbf {\bibinfo {volume} {79}},\ \bibinfo {pages} {161403} (\bibinfo {year} {2009})}\BibitemShut {NoStop}%
\bibitem [{\citenamefont {Gonzalez~Szwacki}(2008)}]{GonzalezSzwacki2008}%
  \BibitemOpen
  \bibfield  {author} {\bibinfo {author} {\bibfnamefont {N.}~\bibnamefont {Gonzalez~Szwacki}},\ }\href {https://doi.org/10.1007/s11671-007-9113-1} {\bibfield  {journal} {\bibinfo  {journal} {Nanoscale Research Letters}\ }\textbf {\bibinfo {volume} {3}},\ \bibinfo {pages} {49} (\bibinfo {year} {2008})}\BibitemShut {NoStop}%
\bibitem [{\citenamefont {Sheng}\ \emph {et~al.}(2009)\citenamefont {Sheng}, \citenamefont {Yan}, \citenamefont {Zheng},\ and\ \citenamefont {Su}}]{Sheng2009}%
  \BibitemOpen
  \bibfield  {author} {\bibinfo {author} {\bibfnamefont {X.-L.}\ \bibnamefont {Sheng}}, \bibinfo {author} {\bibfnamefont {Q.-B.}\ \bibnamefont {Yan}}, \bibinfo {author} {\bibfnamefont {Q.-R.}\ \bibnamefont {Zheng}},\ and\ \bibinfo {author} {\bibfnamefont {G.}~\bibnamefont {Su}},\ }\href {https://doi.org/10.1039/b911519f} {\bibfield  {journal} {\bibinfo  {journal} {Physical Chemistry Chemical Physics}\ }\textbf {\bibinfo {volume} {11}},\ \bibinfo {pages} {9696} (\bibinfo {year} {2009})}\BibitemShut {NoStop}%
\bibitem [{\citenamefont {Choi}\ and\ \citenamefont {et~al.}(2024)}]{Choi2024}%
  \BibitemOpen
  \bibfield  {author} {\bibinfo {author} {\bibfnamefont {H.~W.}\ \bibnamefont {Choi}}\ and\ \bibinfo {author} {\bibnamefont {et~al.}},\ }\bibfield  {journal} {\bibinfo  {journal} {ChemRxiv}\ }\href {https://doi.org/10.26434/chemrxiv-2024-2xnxl} {10.26434/chemrxiv-2024-2xnxl} (\bibinfo {year} {2024})\BibitemShut {NoStop}%
\bibitem [{\citenamefont {Mannix}\ \emph {et~al.}(2015)\citenamefont {Mannix}, \citenamefont {Zhou}, \citenamefont {Kiraly}, \citenamefont {Wood}, \citenamefont {Alducin}, \citenamefont {Myers}, \citenamefont {Liu}, \citenamefont {Fisher}, \citenamefont {Santiago}, \citenamefont {Guest}, \citenamefont {Yacaman}, \citenamefont {Ponce}, \citenamefont {Oganov}, \citenamefont {Hersam},\ and\ \citenamefont {Guisinger}}]{Mannix2015}%
  \BibitemOpen
  \bibfield  {author} {\bibinfo {author} {\bibfnamefont {A.~J.}\ \bibnamefont {Mannix}}, \bibinfo {author} {\bibfnamefont {X.-F.}\ \bibnamefont {Zhou}}, \bibinfo {author} {\bibfnamefont {B.}~\bibnamefont {Kiraly}}, \bibinfo {author} {\bibfnamefont {J.~D.}\ \bibnamefont {Wood}}, \bibinfo {author} {\bibfnamefont {D.}~\bibnamefont {Alducin}}, \bibinfo {author} {\bibfnamefont {B.~D.}\ \bibnamefont {Myers}}, \bibinfo {author} {\bibfnamefont {X.}~\bibnamefont {Liu}}, \bibinfo {author} {\bibfnamefont {B.~L.}\ \bibnamefont {Fisher}}, \bibinfo {author} {\bibfnamefont {U.}~\bibnamefont {Santiago}}, \bibinfo {author} {\bibfnamefont {J.~R.}\ \bibnamefont {Guest}}, \bibinfo {author} {\bibfnamefont {M.~J.}\ \bibnamefont {Yacaman}}, \bibinfo {author} {\bibfnamefont {A.}~\bibnamefont {Ponce}}, \bibinfo {author} {\bibfnamefont {A.~R.}\ \bibnamefont {Oganov}}, \bibinfo {author} {\bibfnamefont {M.~C.}\ \bibnamefont {Hersam}},\ and\ \bibinfo {author} {\bibfnamefont {N.~P.}\ \bibnamefont {Guisinger}},\ }\href
  {https://doi.org/10.1126/science.aad1080} {\bibfield  {journal} {\bibinfo  {journal} {Science}\ }\textbf {\bibinfo {volume} {350}},\ \bibinfo {pages} {1513–1516} (\bibinfo {year} {2015})}\BibitemShut {NoStop}%
\bibitem [{\citenamefont {Feng}\ \emph {et~al.}(2016)\citenamefont {Feng}, \citenamefont {Zhang}, \citenamefont {Zhong}, \citenamefont {Li}, \citenamefont {Li}, \citenamefont {Li}, \citenamefont {Cheng}, \citenamefont {Meng}, \citenamefont {Chen},\ and\ \citenamefont {Wu}}]{Feng2016}%
  \BibitemOpen
  \bibfield  {author} {\bibinfo {author} {\bibfnamefont {B.}~\bibnamefont {Feng}}, \bibinfo {author} {\bibfnamefont {J.}~\bibnamefont {Zhang}}, \bibinfo {author} {\bibfnamefont {Q.}~\bibnamefont {Zhong}}, \bibinfo {author} {\bibfnamefont {W.}~\bibnamefont {Li}}, \bibinfo {author} {\bibfnamefont {S.}~\bibnamefont {Li}}, \bibinfo {author} {\bibfnamefont {H.}~\bibnamefont {Li}}, \bibinfo {author} {\bibfnamefont {P.}~\bibnamefont {Cheng}}, \bibinfo {author} {\bibfnamefont {S.}~\bibnamefont {Meng}}, \bibinfo {author} {\bibfnamefont {L.}~\bibnamefont {Chen}},\ and\ \bibinfo {author} {\bibfnamefont {K.}~\bibnamefont {Wu}},\ }\href {https://doi.org/10.1038/nchem.2491} {\bibfield  {journal} {\bibinfo  {journal} {Nature Chemistry}\ }\textbf {\bibinfo {volume} {8}},\ \bibinfo {pages} {563–568} (\bibinfo {year} {2016})}\BibitemShut {NoStop}%
\bibitem [{\citenamefont {Giannozzi}\ \emph {et~al.}(2009)\citenamefont {Giannozzi}, \citenamefont {Baroni}, \citenamefont {Bonini}, \citenamefont {Calandra}, \citenamefont {Car}, \citenamefont {Cavazzoni}, \citenamefont {Ceresoli}, \citenamefont {Chiarotti}, \citenamefont {Cococcioni}, \citenamefont {Dabo}, \citenamefont {Dal~Corso}, \citenamefont {de~Gironcoli}, \citenamefont {Fabris}, \citenamefont {Fratesi}, \citenamefont {Gebauer}, \citenamefont {Gerstmann}, \citenamefont {Gougoussis}, \citenamefont {Kokalj}, \citenamefont {Lazzeri}, \citenamefont {Martin-Samos}, \citenamefont {Marzari}, \citenamefont {Mauri}, \citenamefont {Mazzarello}, \citenamefont {Paolini}, \citenamefont {Pasquarello}, \citenamefont {Paulatto}, \citenamefont {Sbraccia}, \citenamefont {Scandolo}, \citenamefont {Sclauzero}, \citenamefont {Seitsonen}, \citenamefont {Smogunov}, \citenamefont {Umari},\ and\ \citenamefont {Wentzcovitch}}]{Giannozzi2009}%
  \BibitemOpen
  \bibfield  {author} {\bibinfo {author} {\bibfnamefont {P.}~\bibnamefont {Giannozzi}}, \bibinfo {author} {\bibfnamefont {S.}~\bibnamefont {Baroni}}, \bibinfo {author} {\bibfnamefont {N.}~\bibnamefont {Bonini}}, \bibinfo {author} {\bibfnamefont {M.}~\bibnamefont {Calandra}}, \bibinfo {author} {\bibfnamefont {R.}~\bibnamefont {Car}}, \bibinfo {author} {\bibfnamefont {C.}~\bibnamefont {Cavazzoni}}, \bibinfo {author} {\bibfnamefont {D.}~\bibnamefont {Ceresoli}}, \bibinfo {author} {\bibfnamefont {G.~L.}\ \bibnamefont {Chiarotti}}, \bibinfo {author} {\bibfnamefont {M.}~\bibnamefont {Cococcioni}}, \bibinfo {author} {\bibfnamefont {I.}~\bibnamefont {Dabo}}, \bibinfo {author} {\bibfnamefont {A.}~\bibnamefont {Dal~Corso}}, \bibinfo {author} {\bibfnamefont {S.}~\bibnamefont {de~Gironcoli}}, \bibinfo {author} {\bibfnamefont {S.}~\bibnamefont {Fabris}}, \bibinfo {author} {\bibfnamefont {G.}~\bibnamefont {Fratesi}}, \bibinfo {author} {\bibfnamefont {R.}~\bibnamefont {Gebauer}}, \bibinfo {author} {\bibfnamefont
  {U.}~\bibnamefont {Gerstmann}}, \bibinfo {author} {\bibfnamefont {C.}~\bibnamefont {Gougoussis}}, \bibinfo {author} {\bibfnamefont {A.}~\bibnamefont {Kokalj}}, \bibinfo {author} {\bibfnamefont {M.}~\bibnamefont {Lazzeri}}, \bibinfo {author} {\bibfnamefont {L.}~\bibnamefont {Martin-Samos}}, \bibinfo {author} {\bibfnamefont {N.}~\bibnamefont {Marzari}}, \bibinfo {author} {\bibfnamefont {F.}~\bibnamefont {Mauri}}, \bibinfo {author} {\bibfnamefont {R.}~\bibnamefont {Mazzarello}}, \bibinfo {author} {\bibfnamefont {S.}~\bibnamefont {Paolini}}, \bibinfo {author} {\bibfnamefont {A.}~\bibnamefont {Pasquarello}}, \bibinfo {author} {\bibfnamefont {L.}~\bibnamefont {Paulatto}}, \bibinfo {author} {\bibfnamefont {C.}~\bibnamefont {Sbraccia}}, \bibinfo {author} {\bibfnamefont {S.}~\bibnamefont {Scandolo}}, \bibinfo {author} {\bibfnamefont {G.}~\bibnamefont {Sclauzero}}, \bibinfo {author} {\bibfnamefont {A.~P.}\ \bibnamefont {Seitsonen}}, \bibinfo {author} {\bibfnamefont {A.}~\bibnamefont {Smogunov}}, \bibinfo {author}
  {\bibfnamefont {P.}~\bibnamefont {Umari}},\ and\ \bibinfo {author} {\bibfnamefont {R.~M.}\ \bibnamefont {Wentzcovitch}},\ }\href {https://doi.org/10.1088/0953-8984/21/39/395502} {\bibfield  {journal} {\bibinfo  {journal} {Journal of Physics: Condensed Matter}\ }\textbf {\bibinfo {volume} {21}},\ \bibinfo {pages} {395502} (\bibinfo {year} {2009})}\BibitemShut {NoStop}%
\bibitem [{\citenamefont {van Setten}\ \emph {et~al.}(2018)\citenamefont {van Setten}, \citenamefont {Giantomassi}, \citenamefont {Bousquet}, \citenamefont {Verstraete}, \citenamefont {Hamann}, \citenamefont {Gonze},\ and\ \citenamefont {Rignanese}}]{vanSetten2018}%
  \BibitemOpen
  \bibfield  {author} {\bibinfo {author} {\bibfnamefont {M.}~\bibnamefont {van Setten}}, \bibinfo {author} {\bibfnamefont {M.}~\bibnamefont {Giantomassi}}, \bibinfo {author} {\bibfnamefont {E.}~\bibnamefont {Bousquet}}, \bibinfo {author} {\bibfnamefont {M.}~\bibnamefont {Verstraete}}, \bibinfo {author} {\bibfnamefont {D.}~\bibnamefont {Hamann}}, \bibinfo {author} {\bibfnamefont {X.}~\bibnamefont {Gonze}},\ and\ \bibinfo {author} {\bibfnamefont {G.-M.}\ \bibnamefont {Rignanese}},\ }\href {https://doi.org/10.1016/j.cpc.2018.01.012} {\bibfield  {journal} {\bibinfo  {journal} {Computer Physics Communications}\ }\textbf {\bibinfo {volume} {226}},\ \bibinfo {pages} {39–54} (\bibinfo {year} {2018})}\BibitemShut {NoStop}%
\end{thebibliography}%

\end{document}